\documentstyle[aps,tighten,amssymb]{revtex}

\newcommand{\Christoffel}[2]{{\textstyle\bigl\{\! {#1 \atop #2}\! \bigr\}} }

\begin{document}

\draft

\preprint{UTPT-95-19}

\title{Geodesic and Path Motion in the Nonsymmetric
Gravitational Theory}

\author{J.\ L\'egar\'e and J.\ W.\ Moffat}

\address{Department of Physics, University of Toronto,
Toronto, Ontario, Canada M5S 1A7}

\date{September 21, 1995}

\maketitle

\begin{abstract}
We study the problem of test-particle motion in the Nonsymmetric
Gravitational Theory (NGT) assuming the four-velocity of the particle
is parallel-transported along the trajectory.
The predicted motion is studied on a static, spherically
symmetric background field, with particular attention paid
to radial and circular motions.
Interestingly, it is found that the proper time taken to travel
between any two non-zero radial positions is finite.
It is also found that circular orbits can be supported at lower
radii than in General Relativity for certain forms of motion.

We present three interactions which could be used as alternate methods
for coupling a test-particle to the antisymmetric components of the
NGT field.
One of these takes the form of a Yukawa force in the weak-field limit of a
static, spherically symmetric field,
which could lead to interesting phenomenology.
\end{abstract}


\section{Introduction}

The problem of particle motion in gravitational theory is a long-standing
and complicated one.
Yet, no problem could be more important in verifying the physical
validity of a theory, since we measure a gravitational field by observing
its effect on particle motion.

Several studies have been written on the problem of motion in
General Relativity (GR), both for test-particles and for massive bodies.
Einstein \cite{bib:Einstein} originally postulated that the four-velocity
of a test-particle was parallel-transported along itself during the
course of the particle's motion.
He also showed that such a motion described a path of extremal length.
In 1927, Einstein and Grommer \cite{bib:Einstein_and_Grommer}
showed that the
field equations of GR did not allow for arbitrary motions of mass
concentrations, rendering Einstein's original postulate of geodesic motion
for test-particles spurious.
Beginning in 1938, several workers sought to
derive the equation of particle motion from the field equations,
either by considering
how the field equations governed the motion
of singularities in the gravitational field
\cite{bib:singularities}, or by
investigating the matter-response equation,
$\nabla_\nu T^{\mu\nu}=0$ where $T^{\mu\nu}$ is the stress-energy
tensor, and
its effect on the motion of mass concentrations
\cite{bib:matter_response,bib:Papapetrou}.
In \cite{bib:Wheeler}, Wheeler discusses the potential problems associated
with attempting to derive the equation of motion from the gravitational
field equations.
In particular, he points out that describing concentrations of
mass-energy by singularities in the metric can render a theory
incomplete.
Meanwhile, approaching the problem from the point of view of the
matter-response equation is troublesome, as different stress tensors
yield fundamentally different equations of motion.
As Wheeler points out, however, both of these methods are valid in the
weak-field r\'egime.

The Nonsymmetric Gravitational Theory (NGT) involves a
nonsymmetric, second-rank tensor field $g_{\mu\nu}$,
interpreted as the gravitational potential.
This tensor is refered to as the fundamental tensor.
Although it is tempting to proceed in analogy with GR and associate
with NGT a metric space, it is not clear {\it a priori} how this
structure is to be identified.
In old NGT (mathematically equivalent to Einstein-Straus Unified
Field Theory, but with a different interpretation), it was found
by means of a Cauchy analysis of the field equations
(see \cite{bib:Lichnerowicz}
and~\cite{bib:Maurer-Tison_1,bib:Maurer-Tison_2}; note that
this work was performed in the framework of Einstein-Straus Unified
Field Theory) that the theory contained three possibly different sets
of characteristic surfaces, corresponding to three different metric
structures (see~(30.1)--(30.3) in~\cite{bib:Maurer-Tison_2}).
It is not clear which should be chosen.

In a recent version of NGT, also known as massive or finite-ranged NGT,
the field
equations have been modified so as to correct asymptotic
consistency problems
(see~\cite{bib:DDM,bib:Moffat_NGT_2,bib:Legare_and_Moffat,bib:Clayton}).
In the linearized theory, this corresponds to endowing the antisymmetric
sector of the gravitational field with a mass.
Certain modifications to the compatibility condition have also been made
(see, for instance, (10) of \cite{bib:Legare_and_Moffat}).
These changes are sufficiently significant to cast doubt on
the direct applicability of the precise results of
\cite{bib:Lichnerowicz} and \cite{bib:Maurer-Tison_1,bib:Maurer-Tison_2}.
It is evident, however, that the metric structure of a generic
NGT spacetime still would not be unique.
Current research is exploring the
possibility of generalizing the work described in \cite{bib:Lichnerowicz}
and~\cite{bib:Maurer-Tison_1,bib:Maurer-Tison_2}.

Ignoring for the time being these unanswered questions, we will
assume the metric structure is given by $s_{\mu\nu} \equiv g_{(\mu\nu)}$;
the generalization of most of our results to some other metric structure
amounts essentially to redefining certain metric functions.
We will define $g^{\mu\nu}$ by
$g^{\mu\nu}g_{\mu\sigma}=g^{\nu\mu}g_{\sigma\mu}=\delta^\nu_\sigma$, and
$s^{\mu\nu}$ by $s^{\mu\nu}s_{\mu\sigma}=\delta^\nu_\sigma$.
Note that in the first of these, the order of the indices is important.
We will also define $a_{\mu\nu} \equiv g_{[\mu\nu]}$.
A boldface character will be used to denote a tensor density:
${\bf g}_{\mu\nu} = \sqrt{-g}g_{\mu\nu}$.

The problem of particle motion is no simpler in
NGT than in GR.
Despite its considerable historical value, an analysis of the type
performed by Einstein {\it et al.} in \cite{bib:singularities} is
not desirable due to the interpretational difficulties associated
with introducing singularities into a field theory.
On the other hand, the method of Papapetrou
\cite{bib:matter_response,bib:Papapetrou}
demands {\it a priori} knowledge of
the stress tensor $T^{\mu\nu}$, or at the very least
a physical interpretation of its components.
In GR, the stress tensor is a symmetric tensor; this allows a simple
physical picture of its components as being energy and momentum
densities.
In NGT, however, the stress tensor need not be symmetric.
Although its symmetric components can be interpreted as in GR, is it
unclear how the antisymmetric components should be interpreted.
Furthermore, a discussion analogous to the debate on the uniqueness
of the metric structure outlined above also applies to the stress tensor:
it is not clear which form of the stress tensor
($T^{\mu\nu}$, $T_{\mu\nu}$ or some other combination) should be
interpreted as representing the energy and momentum densities associated
with the gravitational field.
It will nevertheless be informative to investigate the predictions of the
matter-response equation for the unambiguous case of a structureless
test-particle;
this will be done at the end of Section~\ref{sec:conservation_laws}.
The methods of both Einstein \cite{bib:singularities} and Papapetrou
\cite{bib:matter_response,bib:Papapetrou} have been, to one extent
or another, generalized to the case of NGT \cite{bib:generalized_motions}.

Notwithstanding the criticisms of
Wheeler, we will follow Einstein and simply postulate that the
four-velocity of a test-particle is parallel-transported along itself
during its motion.
Section~\ref{sec:parallel_transport} discusses some generalities about
particle motion and parallel-transport; we also introduce the notation
used in the remainder of this paper.
In Section~\ref{sec:conservation_laws}, we derive the first integrals
arising from the equations of motion due to symmetries in the background
spacetime.
Section~\ref{sec:SSS_motion}
applies the results of the previous sections to the so-called geodesic and
path equations for the special case of a static, spherically symmetric
background spacetime.
Radial and circular motions in a static, spherically symmetric field are
discussed in Sections~\ref{sec:radial_motion} and~\ref{sec:circular_motion},
respectively.
In Section~\ref{sec:asymptotic_behaviour}, we briefly demonstrate that the
geodesic and path equations predict similar results in the weak-field
r\'egime;
we derive the first-order correction to the geodesic equation coming from
the path equation, showing that this correction must be consistently
ignored.
Finally, in Section~\ref{sec:modifying_geodesics}, we discuss the possibility
of modifying geodesic motion so as to include other couplings to the
antisymmetric field.

\section{Geodesic and Path Equations}

\label{sec:parallel_transport}

We define a connection $D$ by its effect on a set of basis vectors $e_\mu$:
$D_X e_\mu = C^\beta_{\mu\nu}X^\nu e_\beta$,
where $C^\beta_{\mu\nu}$ are the coefficients of the connection,
which need not be symmetric on $\mu\nu$.
We add to this the requirement that $D_\mu$ obey the Leibnitz rule,
as well as commuting with contractions.
By decomposing vectors onto this basis, we
deduce that the effect of the connection $D$ on vectors
is given in component form by:
$D_\mu v^\lambda = \partial_\mu v^\lambda + C^\lambda_{\rho\mu} v^\rho$.
This establishes the operational use of $D$ on vectors;
the generalization to covectors and tensors of higher-rank is immediate.

A vector $u$ is said to be parallel-transported with respect to a
given curve $x(\tau)$ if
its components at a point $P'$ on the curve are proportional
to its components at another
point $P$ on the curve, an infinitesimal
distance away (see~\cite{bib:Eisenhart}, p.~13):
$D_{\dot x} u=\phi(\tau)u$,
where $\dot x = dx/d\tau$ and $\phi(\tau)$ is an
arbitrary function of $\tau$.
Here, $\tau$ is some parameter for the curve $x(\tau)$.
In particular, if we take $u=\dot x$, we may write in component form:
$u^\nu D_\nu u^\beta = \phi(\tau) u^\beta$.
The term on the right-hand side can be eliminated by
affinely reparametrizing the curve, leading to the
generalized path equation (see also \cite{bib:Regularity}):
\begin{equation}
\label{eq:path_equation}
u^\nu D_\nu u^\beta
= \frac{du^\beta}{d\tau} + C^\beta_{\mu\nu} u^\mu u^\nu = 0
\end{equation}
for some affine parameter $\tau$.
Being nothing more than the definition of parallel-transport of a vector
with respect to a given connection $D$, the generalized path equation is
fundamental in itself; it is not clear that it
could be derived from any more fundamental principle, say, a variational
principle.
The relation of the generalized path equation to paths of extremal
length will be discussed in more detail below.

We will take $u^\beta$ to represent the components of the
particle four-velocity, and we will call
(\ref{eq:path_equation}) the equation of particle motion.

Our definition of the generalized path equation is essentially that
of Eisenhart (see \cite{bib:Eisenhart}, p.~12).
However,
alternate but equivalent definitions of the generalized path equation
exist in the literature.
For instance, Lichnerowicz defines a path $x(\tau)$ of the connection with
coefficients $C^\beta_{\mu\nu}$ by:
\begin{equation}
\label{eq:Lich_path}
\left(\frac{du^\beta}{d\tau}+C^\beta_{\mu\nu} u^\mu u^\nu\right) u^\alpha
- \left(\frac{du^\alpha}{d\tau}+C^\alpha_{\mu\nu} u^\mu u^\nu\right) u^\beta
= 0 ,
\end{equation}
(see~\cite{bib:Lichnerowicz}, pp.~248--249; there,
the generalized path equation
is referred to as a geodesic equation, regardless of the
coefficients $C^\beta_{\mu\nu}$),
where $u=dx/d\tau$ is the tangent vector to the path.
We can recover this definition from the generalized path equation by
multiplying (\ref{eq:path_equation}) by $u^\alpha$ and antisymmetrizing
on $\alpha\beta$.
Tonnelat (see~\cite{bib:Tonnelat}, pp.~145--148)
and Hlavat\'y (see~\cite{bib:Hlavaty}, pp.~64--66) propose similar
definitions.
Although these are equivalent, (\ref{eq:Lich_path}) has the
advantage of being independent of the parametrization of the
curve $x(\tau)$, and holds even for a non-affine parameter.
For our purposes, the added benefit of being able to use a general
parameter is not worth the increase in complexity associated with using
a tensor equation of motion over a vector equation of motion.
We will therefore use (\ref{eq:path_equation}) instead of
(\ref{eq:Lich_path}).

Before proceeding any further, we must specify the coefficients
$C^\beta_{\mu\nu}$.
Although a multitude of connections can be formed in NGT,
there are three that arise more commonly.
The first is the symmetric Levi-Civita connection, whose
coefficients are the usual Christoffel symbols:
\begin{equation}
\label{eq:Christoffel}
\Christoffel{\beta}{\mu\nu}
= \frac{1}{2}s^{\beta\lambda}(\partial_\nu s_{\mu\lambda} +
\partial_\mu s_{\lambda\nu} - \partial_\lambda s_{\mu\nu}) .
\end{equation}
Being compatible with $s_{\mu\nu}$, it is an appealing candidate
for describing test-particle motion, as it leaves the magnitude of
vectors, and hence the four-velocity, unchanged under
parallel transport.
The second is the unconstrained $W$-connection whose torsion
vector $W_\mu$ appears in the
NGT action \cite{bib:Moffat_NGT_2,bib:Legare_and_Moffat,bib:Clayton}.
The third is the $\Gamma$-connection, whose torsion vector
is constrained to vanish.
These last two connections are not independent, but are related by
\begin{equation}
\label{eq:Gamma_W}
\Gamma^\beta_{\mu\nu} = W^\beta_{\mu\nu}
- \frac{2}{3}\delta^\beta_\mu W_\nu ,
\end{equation}
which ensures that the torsion vector $\Gamma_\nu = 0$.
It can be shown (see~\cite{bib:Eisenhart}, pp.~12--13 and pp.~30--31) that
the relation (\ref{eq:Gamma_W}) is sufficient to guarantee
that (\ref{eq:path_equation})
will predict the same motion, regardless of which of these two
connections is used.
This can be seen immediately by introducing (\ref{eq:Gamma_W})
into (\ref{eq:Lich_path}).
The problem is therefore reduced to studying motion predicted by
a Levi-Civita connection and that predicted by a $\Gamma$-connection.

Note that for a non-Levi-Civita connection, the solutions of
(\ref{eq:path_equation}) are not paths of extremal
length \cite{bib:Mann_and_Moffat}.
By contracting (\ref{eq:path_equation}) with $s_{\alpha\beta}$ and
collecting terms, we arrive at
\begin{equation}
\label{eq:EL_with_stuff}
\frac{1}{2}\left(\frac{d}{d\tau}\frac{\partial L}{\partial u^\alpha}
- \frac{\partial L}{\partial x^\alpha}\right)
= s_{\alpha\beta}u^\mu u^\nu\left(\Christoffel{\beta}{\mu\nu}
- C^\beta_{\mu\nu}\right) ,
\end{equation}
where we have introduced $L = s_{\mu\nu}u^\mu u^\nu$ as the magnitude
of the vector $u$.
We recognize the left-hand side of (\ref{eq:EL_with_stuff}) as the
Euler-Lagrange equation for the Lagrangian $L$,
the very Lagrangian used when extremizing the length of a
curve whose tangent vector is $u$.
We conclude that if the right-hand side of (\ref{eq:EL_with_stuff})
vanishes, which is by no means assured, then the length of the
path described by
(\ref{eq:path_equation}) will satisfy the Euler-Lagrange equation and
the path will have extremal length.
This occurs trivially when the connection
appearing in (\ref{eq:path_equation}) is the Levi-Civita connection.
Other connections must be examined on a case-by-case basis.

To be specific, we will refer to the case of (\ref{eq:path_equation})
with the Levi-Civita connection as the geodesic equation; when
using the $\Gamma$-connection, we will refer to this as the
path equation.

\section{Conservation Laws}

\label{sec:conservation_laws}

We will now show that the generalized path equation contains first integrals
of the motion that reflect a symmetry in the background geometry,
allowing us to simplify the description of the motion.
Interestingly, we will see that the first integrals are not always
explicitly guaranteed to exist, leading us to question whether a
conserved quantity is necessarily associated with every symmetry in
the background geometry.

Consider $X_\beta u^\beta$, where $X_\beta = s_{\beta\alpha}X^\alpha$
and where
$u^\beta$ satisfies~(\ref{eq:path_equation});
$X^\beta$ are the components of an arbitrary vector.
Take
\[
\frac{d(X_\beta u^\beta)}{d\tau}
= u^\beta u^\nu D_\nu X_\beta + X_\beta u^\nu D_\nu u^\beta
= u^\beta u^\nu D_{(\nu}X_{\beta)} ;
\]
by direct calculation,
\[
D_{(\nu}X_{\mu)} = \frac{1}{2}\pounds_X[s]_{\mu\nu}
+ X_\beta\left(\Christoffel{\beta}{\mu\nu}-C^\beta_{\mu\nu}\right) ,
\]
where
$\pounds_X[s]_{\mu\nu} = X^\alpha \partial_\alpha s_{\mu\nu}
+ s_{\alpha\nu}\partial_\mu X^\alpha
+ s_{\mu\alpha}\partial_\nu X^\alpha$
is the $\mu\nu$-component of
the Lie derivative of the symmetric part of the fundamental tensor
along a flow
generated by $X$.
Therefore,
\begin{equation}
\label{eq:all_in_one_conservation}
\frac{d(X_\beta u^\beta)}{d\tau}
= \frac{1}{2}\pounds_X[s]_{\mu\nu} u^\mu u^\nu
+ X_\beta u^\mu u^\nu \left(\Christoffel{\beta}{\mu\nu}
-C^\beta_{\mu\nu}\right) .
\end{equation}

If $X$ is a Killing vector, then $\pounds_X[s]=0$ and
(\ref{eq:all_in_one_conservation}) reduces to
\begin{equation}
\label{eq:conserved_momentum}
\frac{d(X_\beta u^\beta)}{d\tau}
= X_\beta u^\mu u^\nu \left(\Christoffel{\beta}{\mu\nu}
-C^\beta_{\mu\nu}\right) .
\end{equation}
If the connection coefficients $C^\beta_{\mu\nu}$ are taken to be Christoffel
symbols, we see that (\ref{eq:conserved_momentum}) is sufficient
to guarantee the existence of a conserved quantity.
Two notable examples are $X=\partial/\partial t$ and
$X=\partial/\partial\phi$, both Killing vectors
of a static, spherically symmetric system.
Writing $s_{tt}=\gamma$ and $s_{\phi\phi}=\beta\sin^2\theta$, we conclude from
(\ref{eq:conserved_momentum}) that
$s_{t\beta}u^\beta=\gamma\dot t\equiv E$ and
$s_{\phi\beta}u^\beta = \beta\dot\phi \sin^2\theta\equiv J$ are
conserved quantities,
corresponding physically to the conservation of energy per unit mass
and angular momentum per unit mass.

The situation is radically different when the $C^\beta_{\mu\nu}$ are
taken to be some connection coefficients other than the Christoffel symbols.
In particular, take $C^\beta_{\mu\nu} = \Gamma^\beta_{\mu\nu}$.
In this case, we see from (\ref{eq:conserved_momentum}) that the fact
that $X$ may
be a Killing vector is not sufficient to guarantee the existence of
a conserved momentum;
we must further require that the relevant connection coefficients
be equal to the corresponding Christoffel symbols.
A case in point is the Killing vector $X=\partial/\partial t$ of a
static, spherically symmetric system.
{}From Appendix~\ref{sec:Wyman_solution}, we see that the only
relevant connection coefficient, $\Gamma^t_{(rt)}$,
is indeed equal to the corresponding Christoffel symbol.
It therefore follows that $s_{t\beta}u^\beta = \gamma\dot t \equiv E$
is a conserved
quantity, again corresponding physically to the conservation of
energy per unit mass.

In contrast, consider the Killing vector
$X=\partial/\partial\phi$ of this same system.
An inspection of the connection coefficients listed in
Appendix~\ref{sec:Wyman_solution} will show that these are not equal to the
corresponding Christoffel symbols.
It follows that $s_{\phi\beta}u^\beta = r^2\dot\phi\sin^2\theta$
is not conserved,
despite the postulated spherical symmetry of the situation.
This might lead one to believe that angular momentum would therefore not be
conserved; in fact, it will be shown in the next section that
there exists a conserved quantity whose asymptotic value is
the conventional angular momentum.

Let $X^\beta = u^\beta$ in (\ref{eq:all_in_one_conservation}), giving
\begin{equation}
\label{eq:conservation_of_mass}
\frac{d(s_{\alpha\beta}u^\alpha u^\beta)}{d\tau}
= 2 s_{\alpha\beta} u^\mu u^\nu u^\alpha \left(\Christoffel{\beta}{\mu\nu}
- C^\beta_{\mu\nu}\right)
= - u^\mu u^\nu u^\alpha D_\alpha s_{\mu\nu} .
\end{equation}
In cases where the right-hand side of (\ref{eq:conservation_of_mass})
vanishes, this yields another constant of the motion: the magnitude of
the velocity.
This will obviously be true for the geodesic equation,
as the Levi-Civita connection is by definition compatible with
the symmetric part of the fundamental tensor.
In other situations, we are faced with the possibility
that the magnitude of the velocity of a particle will not be conserved.
Certain special cases arise when the chosen connection is not that of
Levi-Civita, yet the right-hand side of (\ref{eq:conservation_of_mass})
still vanishes.
This is the case when considering a static, spherically symmetric
background field.
As the asymptotic properties of
such a field in NGT force
$W_\mu=0$ \cite{bib:Clayton,bib:Cornish2}, the right-hand
side of (\ref{eq:conservation_of_mass}) is seen to vanish,
allowing the magnitude of the velocity to be conserved.
However, it must be emphasized that this result is true only for a
static, spherically symmetric field; non-Levi-Civita motion
in a more generic NGT field would most likely
leave the magnitude of the velocity unconserved.

No discussion of conservation laws for particle motion would be complete
without mentioning the matter-response equation, itself a conservation law.
In NGT, the matter-response equation can be written
\cite{bib:Legare_and_Moffat}
\begin{equation}
\label{eq:matter-response}
\partial_\rho {\bf t}^\rho_\lambda - \frac{1}{2}\partial_\lambda g_{\mu\nu}
{\bf T}^{\mu\nu} = 0 ,
\end{equation}
where
\begin{equation}
\label{eq:definition_of_t}
t^\rho_\lambda = \frac{1}{2}(g_{\mu\lambda} T^{\mu\rho}
+ g_{\lambda\mu}T^{\rho\mu}) .
\end{equation}
Given the nonsymmetric nature of both $g_{\mu\nu}$ and $T^{\mu\nu}$, it is
not obvious how (\ref{eq:definition_of_t}) can be inverted to give
$T^{\mu\nu}$ in terms of $t^\mu_\nu$.
It will be sufficient for our needs to note that the relationship between
$t^\mu_\nu$ and $T^{\mu\nu}$ is independent of the properties of
$T^{\mu\nu}$, and thus does not involve any acceleration terms.

If we assume the simplest case of a monopole test-particle, we have
\begin{mathletters}
\label{eq:monopole}
\begin{equation}
\int {\bf T}^{\mu\nu} \, d^3x \ne 0
\end{equation}
\begin{equation}
\int (x^\alpha - X^\alpha) {\bf T}^{\mu\nu} \, d^3x = 0 ,
\end{equation}
\end{mathletters}and similarly for higher-order moments, the integration
being carried out over a hypersurface of constant $t$, as per the
definition of Papapetrou (see \cite{bib:Papapetrou}, p.~154).
Here, $X^\alpha$ is the position of the monopole, while $x^\alpha$ is the
position of the observer.
We take $x^0=X^0=t$ and parametrize the motion using $t$.
It is a simple matter to show that (\ref{eq:monopole}) imply similar
multipole relationships for $t^\mu_\nu$.

Using a method similar in spirit to that of Papapetrou (see
\cite{bib:Papapetrou}, pp.~152--157), we can show that
(\ref{eq:matter-response}) leads to
\begin{equation}
\label{eq:m-r_full_motion}
\frac{d}{dt}\left(\frac{dX^\beta}{dt}\int{\bbox\tau}^{00}\,d^3x\right)
+ \Christoffel{\beta}{\mu\nu}\frac{dX^\mu}{dt}\frac{dX^\nu}{dt}
\int{\bbox\tau}^{00}\,d^3x
= \frac{1}{2}s^{\beta\lambda}\left(\partial_\lambda g_{\mu\nu}
\int{\bf T}^{\mu\nu} \, d^3x - \partial_\lambda s_{\mu\nu}
\frac{dX^\mu}{dt}\frac{dX^\nu}{dt}\int{\bbox\tau}^{00}\,d^3x \right),
\end{equation}
where we define $\tau^{\mu\nu}$ by
$\tau^{\mu\nu}=s^{\lambda\nu}t^\mu_\lambda$.
The first term on the right-hand side of
(\ref{eq:m-r_full_motion})
still contains $T^{\mu\nu}$;
however, given the relationship between $T^{\mu\nu}$ and $\tau^{\mu\nu}$,
we can assert that this term does not contain the acceleration of the
monopole, and that this acceleration is contained only in the first term
on the left-hand side.

If $T^{\mu\nu}$ is taken to be symmetric, we can show that
$\tau^{\mu\nu} = T^{\mu\nu}$; this is the motivation for the definition
of $\tau^{\mu\nu}$.
It then follows that
\[
\int {\bf T}^{\mu\nu}\, d^3x = \frac{dX^\mu}{dt}\frac{dX^\nu}{dt}
\int {\bf T}^{00}\, d^3x ,
\]
which causes the right-hand side
of (\ref{eq:m-r_full_motion}) to vanish.
In such a scenario, we find that (\ref{eq:m-r_full_motion})
reduces to
\[
\frac{d}{ds}\left(m\frac{dX^\beta}{ds}\right)
+ m\Christoffel{\beta}{\mu\nu}\frac{dX^\mu}{ds}\frac{dX^\nu}{ds} = 0 ,
\]
where we have suggestively introduced
\[
m = \frac{ds}{dt}\int{\bbox\tau}^{00}\,d^3x .
\]
Thus (\ref{eq:m-r_full_motion}) reduces to the geodesic equation
when $T^{[\mu\nu]}$ vanishes.
However, note that we have not required that $g_{[\mu\nu]}$ vanish: we are
still working within the realm of NGT.

Of course, $T^{[\mu\nu]} = 0$ is a special case; indeed,
(\ref{eq:m-r_full_motion}) would seem to indicate that
even a simple monopole particle couples directly to
both $T^{[\mu\nu]}$ and $g_{[\mu\nu]}$, and in
NGT there is certainly no need
to force such a contribution to vanish.

The preceeding arguments would certainly seem to lend credence to the
choice of the geodesic equation over the path equation as the correct
description of motion in NGT;
at the very least, (\ref{eq:m-r_full_motion}) is a strong argument for
considering a geodesic equation with a supplementary coupling term on its
right-hand side.
However, the path equation has appeared sufficiently often in the NGT
literature (see, for example,
\cite{bib:Regularity} and \cite{bib:Mann_and_Moffat}, and the
references cited therein) to warrant the
investigation of its physical predictions.
In the remainder of this work, we will describe in parallel some of the
properties of the geodesic and path equations that we have uncovered.
In Section~\ref{sec:modifying_geodesics} below, we will return to the
question outlined above: What happens if $T^{[\mu\nu]}\ne 0$?

\section{Motion in a Static, Spherically Symmetric Field}

\label{sec:SSS_motion}

This section presents the equations and first integrals of
(\ref{eq:path_equation}) in the background geometry of
the static, spherically symmetric
solution described in Appendix~\ref{sec:Wyman_solution}.
These results will be used in the forthcoming sections to study
radial and circular motions.
We consider first the Levi-Civita connection, followed by
the $\Gamma$-connection.

We have already found in Section~\ref{sec:conservation_laws} that
geodesic motion in a static, spherically symmetric field yields as
constants of the motion the energy per unit mass
$E=\gamma\dot t$, and the angular momentum per unit mass
$J=\beta\dot\phi\sin^2\theta$, as well as leaving
$\kappa^2 = s_{\alpha\beta}u^\alpha u^\beta$ unchanged.
The $\theta$-component of the geodesic equation,
\[
\frac{d^2\theta}{d\tau^2}
+ \frac{\beta'}{\beta}\frac{d\theta}{d\tau}\frac{dr}{d\tau}
- \sin\theta\cos\theta \left(\frac{d\phi}{d\tau}\right)^2 = 0 ,
\]
can be satisfied identically by choosing $\theta(\tau_0) = \pi/2$ and
$\dot\theta(\tau_0) = 0$ for some initial proper time $\tau_0$,
corresponding to planar orbits.
Using this, the $r$-component of the geodesic equation becomes
\begin{equation}
\label{eq:geodesic_radial_acceleration}
\frac{d^2r}{d\tau^2}
+ \frac{\alpha'}{2\alpha}\left(\frac{dr}{d\tau}\right)^2
- \frac{\beta'J^2}{2\beta^2\alpha}
+ \frac{\gamma' E^2}{2\alpha\gamma^2} = 0 .
\end{equation}
Although this equation is identical in form to the corresponding equation
in GR,
the predicted motion is different
owing to the different functional content of the metric functions.

Given these constants, the conservation of $\kappa^2$ may be written
\begin{equation}
\label{eq:geodesic_mass}
-\kappa^2 = -s_{\alpha\beta} u^\alpha u^\beta
= \alpha \left(\frac{dr}{d\tau}\right)^2 - \frac{E^2}{\gamma}
+ \frac{J^2}{\beta} .
\end{equation}

Results similar to those listed above can also be derived for the
path equation, the only difficulty arising from the slightly more
complex nature of the connection coefficients.

The $t$-component of the path equation was integrated
in Section~\ref{sec:conservation_laws}, and found to lead to a constant
of the motion: $E=\gamma\dot t$.
This constant can be interpreted physically by examining its value
asymptotically in the flat-space region, where
it is found to correspond to the particle energy per unit mass.

We have also discussed in Section~\ref{sec:conservation_laws} how,
barring unusual circumstances,
the path equation will in general not leave the magnitude of the
velocity, $\kappa^2$, unchanged.
However, in certain special cases, of which the static, spherically symmetric
field is one, a conserved $\kappa^2$
can be extracted from the path equation.
This is by virtue of the form of (\ref{eq:conservation_of_mass}), whose
rightmost member precludes any contribution from the antisymmetric
piece of $D_\alpha s_{\mu\nu}$.
It can be shown that, in the static, spherically symmetric solution to the
NGT field equations, the only contribution coming from this covariant
derivative is indeed antisymmetric on its indices.
We therefore have a conserved $\kappa^2$:
$\kappa^2 = s_{\alpha\beta}u^\alpha u^\beta$.

As was the case for the geodesic equation,
the $\theta$-component of the path equation,
\[
\frac{d^2\theta}{d\tau^2}
+ \frac{\beta\beta'+ff'}{\beta^2+f^2}\frac{dr}{d\tau}\frac{d\theta}{d\tau}
- \sin\theta\cos\theta\left(\frac{d\phi}{d\tau}\right)^2 = 0 ,
\]
can be satisfied identically by choosing $\theta(\tau_0)=\pi/2$ and
$\dot\theta(\tau_0)=0$, so that orbits will lie in a plane.
Here, $f$ is defined by $f\sin\theta=g_{[\theta\phi]}$.
The $\phi$-component of the path equation then leads to
\[
\frac{d^2\phi}{d\tau^2} + \frac{\beta\beta'+ff'}{(\beta^2+f^2)}
\frac{dr}{d\tau}\frac{d\phi}{d\tau}
= \frac{d\ln(\sqrt{\beta^2+f^2}\dot\phi/J)}{d\tau} = 0 ,
\]
where $J$ is a constant having dimensions of a length.
Solving for $J$, we have
\[
J = \sqrt{\beta^2+f^2}\frac{d\phi}{d\tau} .
\]
The physical significance of $J$ is found by taking the limit
$r\rightarrow\infty$.
{}From (\ref{eq:app_larger_r}), we see that $f\rightarrow 0$ exponentially,
so that $J$ behaves asymptotically as $r^2\dot\phi$,
which we identify with the conventional
angular momentum (per unit mass).
We therefore conclude that it is $J=\sqrt{\beta^2+f^2}\dot\phi$ and
not $\beta\dot\phi$ which represents the axial component of the
angular momentum of the path equation.
Near the origin, where $f$ is not small with respect to $\beta$, the
value of $J$ for a given $\dot\phi$ will be very different from
the conventional angular momentum.

Having determined the expression for the constant $J$, we can use this
to give an expression for $\kappa^2$:
\begin{equation}
\label{eq:path_mass}
- \kappa^2 = - s_{\alpha\beta}u^\alpha u^\beta
= \alpha\left(\frac{dr}{d\tau}\right)^2
- \frac{E^2}{\gamma} + \frac{J^2\beta}{\beta^2+f^2}
\end{equation}

Use of these constants of the motion allows us to simplify
the $r$-component of the path equation:
\begin{equation}
\label{eq:path_radial_acceleration}
\frac{d^2r}{d\tau^2}
+ \frac{\alpha'}{2\alpha}\left(\frac{dr}{d\tau}\right)^2
+ \frac{\gamma'E^2}{2\alpha\gamma^2}
- \frac{J^2}{2}
\frac{\beta'(\beta^2-f^2)+2ff'\beta}{\alpha(\beta^2+f^2)^2} = 0 .
\end{equation}

We can combine path and geodesic motion by introducing the function
$\psi(r)$, given by
$\psi=\psi_g=1/\beta$ for the geodesic equation, and
$\psi=\psi_p=\beta/(\beta^2+f^2)$ for the path equation.
We then have
\begin{equation}
\label{eq:both_radial_acceleration}
\frac{d^2r}{d\tau^2}
+ \frac{\alpha'}{2\alpha}\left(\frac{dr}{d\tau}\right)^2
+ \frac{\gamma'E^2}{2\alpha\gamma^2}
- \frac{J^2\psi'}{2\alpha} = 0
\end{equation}
instead of (\ref{eq:geodesic_radial_acceleration}) and
(\ref{eq:path_radial_acceleration}), and
\begin{equation}
\label{eq:both_mass}
-\kappa^2 = \alpha\left(\frac{dr}{d\tau}\right)^2
- \frac{E^2}{\gamma} + J^2\psi
\end{equation}
instead of
(\ref{eq:geodesic_mass}) and (\ref{eq:path_mass}).

We can get a simple comparison of the geodesic and path equations by
first expressing (\ref{eq:both_radial_acceleration}) in terms of
$r(t)$:
\[
\frac{d^2r}{dt^2} + \frac{\alpha'}{2\alpha}\left(\frac{dr}{dt}\right)^2
+ \frac{\gamma'}{2\alpha} - \frac{\beta}{2\alpha}\frac{d\ln(\psi)}{dr}
\left(\frac{d\phi}{dt}\right)^2 = 0 .
\]
The initial conditions for this differential equation are $t=t_0$,
$r_0=r(t_0)$,
$\dot r_0=\dot r(t_0)$, and $\dot\phi_0=\dot\phi(t_0)$, where a dot represents
differentiation with respect to $t$; note that we do not need to specify
$\phi_0=\phi(t_0)$ as it does not appear in the differential equation.
{}From the constancy of $J$ and $E$, we have that
\[
\dot\phi^2 = \frac{\psi/\psi_0}{\beta/\beta_0}\dot\phi_0{}^2 ,
\]
where we have adopted the notation that
a subscript ``$0$'' on a quantity indicates that the quantity is to be
evaluated using the initial data.
It follows that the initial acceleration of the particle is
\[
\left.\frac{d^2r}{dt^2}\right|_0 = -\frac{\alpha_0'\dot r_0{}^2}{2\alpha_0}
- \frac{\gamma_0'}{2\alpha_0}
+ \frac{\beta_0\dot\phi_0{}^2}{2\alpha_0}
\left.\frac{d\ln(\psi)}{dr}\right|_0 .
\]
The rightmost term distinguishes the geodesic and path equations.
It is always true that $\ln(\psi_g/\psi_p) \ge 0$, and for all $r \ge M$
we have $\ln(M^2\psi_g) \le 0$;
therefore $\ln(M^2\psi_g) \le \ln(M^2\psi_p) \le 0$.
In the large $r$ limit, $\ln(\psi_p)\approx\ln(\psi_g)$.
If we assume that $f(r)$ does not behave pathologically anywhere between
$r=M$ and $r\rightarrow +\infty$, it then follows that
\[
\frac{d\ln(\psi_g)}{dr} \le \frac{d\ln(\psi_p)}{dr} ,
\]
with $d\ln(\psi_p)/dr \le 0$.
This allows us to conclude that initially,
given identical initial conditions, a
particle obeying the path equation would feel a greater inward
radial acceleration
than a particle obeying the geodesic equation.
We could imagine a thought experiment where two particles,
one obeying the path equation and the other the geodesic equation,
are weakly deflected by some gravitational source.
Given our previous analysis, we would expect that the first particle,
in obeying the path equation, would be deflected through a larger angle
than the second particle.
In fact, although such a calculation would take us too far off course,
an interesting method to determine experimentally whether
an NGT test-particle
obeys the path equation or the geodesic equation would be to set up just
such a scattering experiment (see, for instance, \cite{bib:Goldstein},
pp.~105--119).
Given certain values for the parameters, one might be able to determine
a bound on the scattering angle with sufficient precision to distinguish
between the geodesic and path equations.

We will now use the results of this section to study the properties of
radial and circular test-particle motion in a static, spherically-symmetric
NGT background.

\section{Radial Motion}

\label{sec:radial_motion}

It is interesting to note that the geodesic and path equations differ
only in their angular content: if $J=0$, they predict the same motion.
This can be seen in both (\ref{eq:both_radial_acceleration})
and (\ref{eq:both_mass}).
It follows that a study of the radial motion of a particle is,
in fact, a generic result, independent of which connection was used to
study the motion.

A particularly convenient form of the equation of motion is found by
introducing reduced variables into
(\ref{eq:both_mass}):
\begin{equation}
\label{eq:path_energy_equation}
\gamma \alpha \left(\frac{dr}{d\tilde\tau}\right)^2
+ \gamma(1+\tilde J^2\psi)
= \tilde K + \tilde V = \tilde E^2 ,
\end{equation}
where $\tilde E=E/\kappa$, $\tilde J=J/\kappa$ and $\tilde\tau=\kappa\tau$.
$\tilde K=\gamma\alpha(dr/d\tilde\tau)^2$ and
$\tilde V=\gamma(1+\tilde J^2\psi)$ are the reduced kinetic and potential
energies, respectively.

{}From (\ref{eq:path_energy_equation}),
we can compute the acceleration of a radially
infalling particle ($\tilde J=0$):
\[
\frac{d}{d\tilde\tau}\left(\frac{dr}{d\tilde\tau}\right)^2
= 2\frac{dr}{d\tilde\tau}\frac{d^2r}{d\tilde\tau^2}
= - \frac{\tilde E^2}{\alpha\gamma}\frac{d(\alpha\gamma)}{dr}
\frac{dr}{d\tilde\tau} + \frac{1}{\alpha^2}\frac{d\alpha}{dr}
\frac{dr}{d\tilde\tau} .
\]
Dividing through by $dr/d\tilde\tau$ gives
\[
\left.\frac{d^2r}{d\tilde\tau^2}\right|_{\rm in}
= \frac{1}{2\alpha^2}\left(\frac{d\alpha}{dr}
- \frac{\tilde E^2}{\gamma^2}\frac{d(\alpha\gamma)}{dr}\right) .
\]
This expression is best evaluated in the $\nu$-coordinate system
(the $\nu$-coordinate system is described
in Appendix~\ref{sec:Wyman_solution}).
It is found that
\begin{equation}
\label{eq:path_acceleration_in_falling}
\left.\frac{d^2\nu}{d\tilde\tau^2}\right|_{\rm in}
= \frac{1}{2\xi(\nu)}\left[
\frac{2(a\sinh(a\nu)+b\sin(b\nu))}{\cosh(a\nu)-\cos(b\nu)}
(\tilde E^2-e^\nu)-e^\nu\right] ,
\end{equation}
where
\[
\xi(\nu) = \frac{M^2 (1+s^2)}{(\cosh(a\nu)-\cos(b\nu))^2} .
\]
Since $\nu \le 0$, we see that the particle is always attracted toward
the larger negative values of $\nu$.

The acceleration of a static particle is more easily found from
(\ref{eq:path_radial_acceleration}) or
(\ref{eq:both_radial_acceleration}).
Dividing this through by $\kappa^2$ and setting $dr/d\tilde\tau=0$ and
$\tilde J=0$ gives
\begin{equation}
\label{eq:path_acceleration_static}
\left.\frac{d^2r}{d\tilde\tau^2}\right|_{\rm st}
= - \frac{\gamma'\tilde E^2}{2\alpha\gamma^2} .
\end{equation}
Again, the $\nu$-coordinate system is more useful for
evaluating this expression, giving
\begin{equation}
\label{eq:path_acceleration_static_2}
\left.\frac{d^2\nu}{d\tilde\tau^2}\right|_{\rm st}
= -\frac{\tilde E^2}{2\xi(\nu)} .
\end{equation}

To work out the turning points, we consider
(\ref{eq:path_energy_equation})
in the $\nu$-coordinate system;
we find that there is only one turning point, at
\begin{equation}
\label{eq:turning_points}
\nu_t = \ln(\tilde E^2)
= \ln(\xi(\nu_i) \dot\nu_i{}^2+e^{\nu_i}) ,
\end{equation}
where $\nu_i$ is the initial radial position and
$\dot\nu_i$ is the initial radial velocity.
In order to ensure that the particle velocity is real, we must
have $\nu \le \nu_t$.
However, since there is only one turning point,
once the particle begins to head towards $\nu\rightarrow -\infty$, it
continues to do so, barring the application of some other force.
It is therefore informative to compute the proper time required for
the particle to reach some $\nu_f$
after starting from $\nu_i$.
This is found by inverting
(\ref{eq:path_energy_equation}) in the $\nu$-coordinate system:
\[
\tilde\tau = (M^2(1+s^2))^{1/2}\int_{\nu_i}^{\nu_f}
\frac{d\nu}{(\cosh(a\nu)-\cos(b\nu))(\tilde E^2-e^\nu)^{1/2}} .
\]
We consider the case $\nu_f < \nu_i \le 0$.
Bounding $\tilde E^2-e^\nu$ by
$\tilde E^2-e^{\nu_i}=\xi(\nu_i)\dot\nu_i{}^2$ and
$\cos(b\nu)$ by $1$, we have that
\begin{equation}
\label{eq:bound_in_time}
\tilde\tau \le \left(\frac{M^2(1+s^2)}{a^2\xi(\nu_i)
\dot\nu_i{}^2}\right)^{1/2}
(\coth(a\nu_i/2) - \coth(a\nu_f/2)) .
\end{equation}
This is a finite number for all $\nu_i$ and $\nu_f$, with
the exception
$\nu_i\rightarrow 0$, where $\tilde\tau\rightarrow+\infty$.
We see that a particle can travel between any two non-zero $\nu$ values
in a finite amount of proper time.

There is a further complication associated with the physical interpretation
of the results we have arrived at: it can been seen from
(\ref{eq:beta_in_nu}) that when $\nu$ goes from $\nu > \nu_0$ to
$\nu < \nu_0$, the metric undergoes a signature change, with $\beta(\nu)$
going from being positive to being negative.
At the point $\nu=\nu_0$, the metric is necessarily degenerate.
However we have found that a particle, regardless of
whether it obeys the geodesic
equation or the path equation, proceeds through the point $\nu=\nu_0$
and beyond without hindrance, despite the fact that at $\nu=\nu_0$,
it would appear that
the fundamental assumption of metric-non-degeneracy,
so crucial to the postulates of the theory, is no longer valid.

Although these results are reminiscent of the Schwarzschild
singularity in GR
(see \cite{bib:MTW}, p.~663), it should be kept in mind that we
are dealing here with a vacuum solution.
It has yet to be established whether the Wyman solution is actually
the final state of some generic stellar collapse problem
(see~\cite{bib:collapse}).

\section{Circular Motion}

\label{sec:circular_motion}

It is interesting to compare the predictions of the geodesic and path
equations in the special case of circular motion as
this displays the most striking difference between these two
forms of motion: their ability or inability to support circular
orbits for certain values of the parameters.

In order to explore the existence of circular orbits,
we minimize $\tilde V$ with respect to $r$:
\[
0 = \frac{d\tilde V}{dr}
= \frac{d\gamma}{dr}+ \tilde J^2\frac{d(\gamma\psi)}{dr} .
\]
Solving for $\tilde J$ gives
\begin{equation}
\label{eq:path_angular_momentum}
\tilde J^2 = - \frac{d\gamma/dr}{d(\gamma\psi)/dr} .
\end{equation}
The location of the minimum of $\tilde V$ is the radius of
a circular orbit for a particular value of $\tilde J$.
Typically in GR, the approach is to solve (\ref{eq:path_angular_momentum})
for $r$ as a function of $\tilde J$.
Analysis of this equation shows that there is a minimum
allowable value of the angular momentum, $\tilde J_0$.
However, this approach is not practical in NGT, as the resulting
equations are seemingly intractable.
In turns out to be more profitable to turn the problem around and
to view (\ref{eq:path_angular_momentum}) as giving the value of
$\tilde J$ necessary to establish an orbit of a given radius.
It is also useful to work in the $\nu$-coordinate system, as
this allows us to investigate the region near $r=M$.
We can use (\ref{eq:really_obvious_relation}) to convert
$\nu$ values to conventional radial values.

$\tilde J$ is a one-to-many function of $\nu$.
In particular, in the region $\nu>\nu_0$, $\tilde J$ is a one-to-two
function of $\nu$.
There must therefore be a local extremum, which turns out to be
a minimum.
This local minimum will
correspond to the smallest allowable angular momentum.
We find this local minimum by solving
\begin{equation}
\label{eq:path_local_minimum}
\frac{d\tilde J^2}{d\nu}
= - \frac{d}{d\nu}\frac{d\gamma/d\nu}{d(\gamma\psi)/d\nu} = 0
\end{equation}
for $\nu=\nu_m$, and then evaluating
$r_m{}^2 = \beta(\nu_m)$ and
$\tilde J^2(\nu_m)= \tilde J_m{}^2$.
In GR, the resulting values (in conventional spherical coordinates)
are $r_m = 6M$ and
$\tilde J_m{}^2 = 12 M^2$ (see \cite{bib:MTW}, p.~662)
Unfortunately, (\ref{eq:path_local_minimum}) cannot be solved in closed form
when $\gamma(\nu)$ and $\psi(\nu)$ take on their NGT form
(\ref{eq:exact_Wyman}).
However, in formulating the problem in this fashion, it becomes
straightforward to solve (\ref{eq:path_local_minimum}) numerically for
$\nu_m$.
As an example, we take $s=0.9$ and $\mu M=0$.
Typically, a more reasonable choice
might be to take $\mu M \sim 10^{-9}$ or so;
however, since we are concerned
with values of $\nu$ near the origin, the problems generated
in NGT by having $\mu M=0$ are not relevant here
(see~\cite{bib:DDM} for a discussion of the problems associated with
having $\mu M=0$), and
taking $\mu M=0$ simply represents an extreme case.
The value $s=0.9$ is chosen so as to emphasize the observed behaviour.
Numerically, it is found that for the
geodesic equation, $\nu_m \approx -0.40602$,
$r_m \approx 5.9930 M$, and
$\tilde J_m{}^2 \approx 11.995 M^2$, while for the
path equation, $\nu_m \approx -0.40511$,
$r_m \approx 6.0040 M$, and
$\tilde J_m{}^2 \approx 12.003 M^2$.
Generically, the geodesic equation yields smaller
values than in GR, while the path equation yields values
larger than the corresponding GR values.

The advantage of using the $\nu$-coordinate system is that the
statements of the previous paragraphs are not tied down to some
expansion, valid only in a limited region of spacetime; results
calculated using the $\nu$-coordinate system can be considered exact and
valid for all $\nu$ in the case of infinite-ranged NGT, or approximately
true for finite-ranged NGT, in those regions where $\mu r \ll 1$ and when
$\mu M$ can be considered small.
In particular, this is the case near the origin, where $r \sim M \ll 1/\mu$,
and we may conclude that
the path equation does not support circular orbits lying below $r=2M$.
On the other hand, circular orbits for the geodesic equation
can be made to approach the origin arbitrarily closely by selecting
successively larger values of the parameter $s$.

Since the exact Wyman solution is not a solution to the field equations
of massive NGT,
one may question the wisdom of using it to
investigate particle motion in NGT, not to mention
the preceding treatment of circular orbits.
Indeed, if applicable at all, the Wyman solution can only be used
near the origin, where we would expect the range $1/\mu$ of
the antisymmetric components of the field to have little effect.
The study of circular orbits provides an excellent opportunity to show
that using the Wyman solution gives qualitatively similar results to using
an asymptotic solution to the field equations.
In fact, if we repeat the preceding analysis of circular orbits in
the path equation, but use (\ref{eq:app_larger_r}) instead of the
Wyman solution, we find that the minimum circular orbit for $s=0.9$
occurs at $r_m \approx 6.0033 M$, with
$J_m{}^2 \approx 12.003 M^2$.
The geodesic equation cannot be similarly verified, as its minimum
circular orbits lie below $r=2M$, where (\ref{eq:app_larger_r}) can
no longer be trusted.

\section{Behaviour in the Asymptotic Regions of Spacetime}

\label{sec:asymptotic_behaviour}

By studying the geodesic and path equations
in a static, spherically symmetric field, we were able to see
in the previous two sections that there are
differences in the resulting motion in the strong-field r\'egime, in
particular for the case of circular motion.
In this brief section, we will demonstrate explicitly for the case of
a static, spherically symmetric background field that the difference
between the geodesic and path equations in the weak-field
r\'egime is of higher-order, and can safely be neglected.

{}From (\ref{eq:app_larger_r}), we know that $f(r)<r^2$ for $r>M$.
This behaviour is emphasized at larger $r$, as $f(r)$ is damped
exponentially.
Given this behaviour, we will consider the difference between the path
equation and the geodesic equation by expanding in powers of
$\varepsilon = f(r)/r^2$;
we will show that this difference must be neglected in order
to be consistent with our other approximations.

Take (\ref{eq:path_equation}) with
$C^\beta_{\mu\nu}=\Gamma^\beta_{\mu\nu}$, and rewrite this as
\begin{equation}
\label{eq:difference}
\frac{du^\beta}{d\tau} + \Christoffel{\beta}{\mu\nu} u^\mu u^\nu
= \left(\Christoffel{\beta}{\mu\nu}
- \Gamma^\beta_{\mu\nu}\right) u^\mu u^\nu
\end{equation}
by adding and subtracting a Christoffel symbol.

Recall that the $t$- and $\theta$-components had the same behaviour in
both the geodesic and path equations.
We need only consider the $\phi$- and $r$-components.
Since the $\phi$-component is immediately integrated, its expansion
is trivial:
\[
\frac{d\phi}{d\tau} \approx \frac{J^2}{r^4}\left(1-\frac{f^2}{r^4}\right).
\]
where $J$ represents the value of the angular momentum in both the
geodesic and
path equations.
Expanding the right-hand side of the radial
component of (\ref{eq:difference})
to lowest order in $f(r)/r^2$, we arrive
at
\begin{equation}
\label{eq:difference_lowest_order}
\frac{d^2r}{d\tau^2} + \frac{\alpha'}{2\alpha}\left(\frac{dr}{d\tau}\right)^2
+ \frac{\gamma'E^2}{2\alpha\gamma^2} - \frac{J^2}{\alpha r^3}
= \left(\frac{ff'}{r^6}-\frac{3f^2}{r^7}\right)\frac{J^2}{\alpha} .
\end{equation}
The left-hand side of this is nothing more than the geodesic equation.
The right-hand side is of order $\varepsilon^2$;
approximating $f(r)$ and $f'(r)$ by their maximum values of
$f(r) \sim sM^2/3$ and $f'(r) \sim -s\mu M^2/3$,
we could conclude that, for large $r$, the difference between the path
equation and the geodesic equation is of order $1/r^6$.
Adding to this the exponentially-damped behaviour of
$f(r)$, we could conclude that for any $r>2M$ where
(\ref{eq:app_larger_r}) can be trusted,
the geodesic and path equations will yield identical results.
Much stronger than this, however, is the fact that the corrections to
$\gamma(r)$ and $\alpha(r)$ in (\ref{eq:app_larger_r}) are themselves
of order
$\varepsilon^2$ (see (4.6) in \cite{bib:Clayton}, and the
discussion preceding it).
Thus, if we accept (\ref{eq:app_larger_r}) as valid approximations to
some exact solution of the NGT field equations, in order to be consistent
with this approximation, we must drop the right-hand side of
(\ref{eq:difference_lowest_order}).

This motivates the use of the geodesic equation in weak-field
regions of spacetime.

\section{Modifying Geodesic Motion}

\label{sec:modifying_geodesics}

In the previous section, it was shown that path motion
converges very rapidly to geodesic motion.
However, by its very nature geodesic motion does not couple the
particle directly to the antisymmetric components of the NGT field.
Moreover, we saw at the end of Section~\ref{sec:conservation_laws} that
even a monopole test-particle must, in the most general case,
couple to $a_{\mu\nu}$.
In this section, we present three possible couplings.
For simplicity, these are linear in the particle velocity.

Since we are dealing in this section only with the weak-field
regions of spacetime, we will consider modifying only the geodesic
equation; from the previous section, we know that the path equation will
yield similar results in these regions.

The couplings will be derived from a scalar Lagrangian $L$, which we
take to have the form $L=(1/2) A_\mu u^\mu$, where $A$ is a
covector independent of the particle velocity.
The Euler-Lagrange equation of particle mechanics allows us to conclude
that the contribution of such a Lagrangian to the equation of motion will
be of the form
\[
\frac{d}{d\tau}\frac{\partial L}{\partial u^\alpha}
- \frac{\partial L}{\partial x^\alpha}
= - f_{[\alpha\mu]}u^\mu ,
\]
where $f_{[\alpha\mu]} = \partial_{[\alpha}A_{\mu]}$.

Let $L=L_1 + L_2 + L_3$, where
\begin{mathletters}
\label{eq:NGT_Lagrangians}
\begin{eqnarray}
L_1 &=&
\label{eq:dual_coupling}
\frac{1}{2} \lambda_1 {}^*\!F^\eta u_\eta =
\frac{\lambda_1}{2}
\frac{\epsilon^{\eta\mu\nu\lambda}}{\sqrt{-g}}
F_{[\mu\nu\lambda]}s_{\eta\sigma} u^\sigma \\
L_2 &=&
\label{eq:direct_coupling}
\frac{1}{2} \lambda_2 g^{[\mu\nu]}
F_{[\mu\nu\lambda]} u^\lambda \\
L_3 &=&
\label{eq:skew_divergence}
\frac{\lambda_3}{2}
\frac{\partial_\nu{\bf g}^{[\lambda\nu]}}{\sqrt{-g}}
s_{\lambda\mu} u^\mu .
\end{eqnarray}
\end{mathletters}The components of the
field-strength tensor $F_{[\mu\nu\lambda]}$ are defined by
\begin{equation}
\label{eq:curl_skew_g}
F_{[\mu\nu\lambda]}
= \partial_{[\lambda}g_{\mu\nu]}
= \frac{1}{3}(\partial_\lambda a_{\mu\nu}
+ \partial_\mu a_{\nu\lambda}
+ \partial_\nu a_{\lambda\mu}) .
\end{equation}
The constants $\lambda_i$ ($i=1,2,3$) couple the test-particle to the NGT
skew field, and have dimensions of a length.
The symbol $\epsilon^{\mu\nu\lambda\eta}$ is the fully antisymmetric
Levi-Civita tensor density, defined by
\[
\epsilon^{\mu\nu\lambda\eta} = \left\{
\begin{array}{cl}
+1 & \mbox{if $\mu\nu\lambda\eta$ is an even permutation of $0123$,} \\
-1 & \mbox{if $\mu\nu\lambda\eta$ is an odd permutation of $0123$,} \\
0 & \mbox{otherwise.}
\end{array}
\right.
\]

It will be found below that only $L_1$ will generate any interaction
at all in a static, spherically symmetric field.
This does not, of course, exclude the possibility that $L_2$ and $L_3$
could more accurately reflect the desired coupling between the test-particle
and the antisymmetric components of the NGT field.
It does, however, make their contribution
trivial in our particular case of interest.
Although for the sake of completeness we will derive the
additional terms in the
equation of motion for the case when all three interactions are
present, we will in the end ignore $L_2$ and $L_3$, knowing that they
will in no way affect our results.

Using the Euler-Lagrange equation, we find that
\begin{equation}
\label{eq:skew_force}
f_{[\alpha\mu]}
= \lambda_1 \partial_{[\alpha}
\left(\frac{\epsilon^{[\eta\sigma\nu\lambda]}}{\sqrt{-g}}
F_{[\sigma\nu\lambda]} s_{\mu]\eta}\right)
+ \lambda_2 \partial_{[\alpha}
\left(g^{[\eta\nu]}F_{[\eta\nu\mu]]}\right)
+ \lambda_3 \partial_{[\alpha}\left(
\frac{\partial_\nu{\bf g}^{[\eta\nu]}}
{\sqrt{-g}} s_{\mu]\eta}\right) ,
\end{equation}
such that the equation of motion is now written
\begin{equation}
\label{eq:modified_geodesic}
\frac{du^\beta}{d\tau}
+ \Christoffel{\beta}{\mu\nu} u^\mu u^\nu
= \kappa^2 s^{\beta\alpha}f_{[\alpha\mu]} u^\mu ,
\end{equation}
where $\kappa^2 = s_{\mu\nu} u^\mu u^\nu$ is the magnitude of the velocity.

As we will only be interested in the behaviour of
(\ref{eq:modified_geodesic}) in a static, spherically symmetric field,
it will be of no interest to re-derive the conservation laws
discussed in Section~\ref{sec:conservation_laws}.
Suffice it to say that the techniques of that section can easily
be generalized, and the first integrals are found to be slightly
different.
It will be simpler for us to derive the constants of the motion
``by hand'' below.
It is straightforward to show that $\kappa^2$ is conserved.
The appearance of the
magnitude of the velocity in the equation of motion in this
fashion has an important implication: for a massless particle (such as
a photon), $\kappa^2=0$ and the right-hand
side of (\ref{eq:modified_geodesic}) is seen to vanish.
Therefore, in this scheme, massless particles will not couple to the
antisymmetric field.

The case $\kappa^2=0$ corresponds to pure geodesic motion, which was
discussed in previous sections.
We will therefore concentrate on the case $\kappa^2\ne 0$.
Since $\kappa^2$ is conserved,
we can scale the proper time $\tau$ in such a
way that $\kappa^2=1$; henceforth, we assume that this has been done.

In the static, spherically symmetric field, the skew field-strength
tensor $F_{[\mu\nu\lambda]}$ has only one independent, non-zero
component,
\[
F_{[\theta\phi r]} = \frac{1}{3}\partial_r a_{\theta\phi}
= \frac{1}{3} f'\sin\theta .
\]
On the other hand, because of the chosen form of the fundamental tensor,
$\partial_\nu{\bf g}^{[\lambda\nu]}$ can be shown to vanish.
{}From (\ref{eq:skew_force}), it follows that the tensor
$f_{[\alpha\sigma]}$ also has
one independent component:
\[
f_{[rt]}
= \frac{d}{dr}
\left(\frac{\lambda_1\gamma f'}{\sqrt{\alpha\gamma(r^4+f^2)}}\right) .
\]
For convenience, we will write $\lambda_1=\lambda$, since the $\lambda_2$
and $\lambda_3$ terms in (\ref{eq:skew_force}) make no contribution to
the equation of motion.

Using (\ref{eq:modified_geodesic}), we find that the equations of
motion for a particle in the field of
a static, spherically symmetric source are
given by
\begin{mathletters}
\begin{eqnarray}
0 &=&
\label{eq:t_component}
\frac{d^2t}{d\tau^2}
+ \frac{\gamma'}{\gamma}\frac{dt}{d\tau}\frac{dr}{d\tau}
+ \frac{\lambda}{\gamma}\frac{dr}{d\tau}\frac{d}{dr}
\left(\frac{\gamma f'}{\sqrt{\alpha\gamma(r^4+f^2)}}\right) \\
0 &=&
\label{eq:r_component}
\frac{d^2r}{d\tau^2}
+ \frac{\alpha'}{2\alpha}\left(\frac{dr}{d\tau}\right)^2
- \frac{r\sin^2\theta}{\alpha}\left(\frac{d\phi}{d\tau}\right)^2
- \frac{r}{\alpha}\left(\frac{d\theta}{d\tau}\right)^2
+ \frac{\gamma'}{2\alpha}\left(\frac{dt}{d\tau}\right)^2
+ \frac{\lambda}{\alpha}\frac{dt}{d\tau}\frac{d}{dr}
\left(\frac{\gamma f'}{\sqrt{\alpha\gamma(r^4+f^2)}}\right) \\
0 &=&
\label{eq:theta_component}
\frac{d^2\theta}{d\tau^2}
+ \frac{2}{r}\frac{d\theta}{d\tau}\frac{dr}{d\tau}
- \sin\theta\cos\theta\left(\frac{d\phi}{d\tau}\right)^2 . \\
0 &=&
\label{eq:phi_component}
\frac{d^2\phi}{d\tau^2} + \frac{2}{r}\frac{d\phi}{d\tau}\frac{dr}{d\tau}
+ 2\frac{\cos\theta}{\sin\theta}\frac{d\phi}{d\tau}\frac{d\theta}{d\tau} .
\end{eqnarray}
\end{mathletters}

We can satisfy (\ref{eq:theta_component}) identically by letting
$\theta(\tau_0)=\pi/2$ and $\dot\theta(\tau_0)=0$ for some initial
proper time $\tau_0$.
It follows that orbits lie in a plane.
Meanwhile, we conclude
from (\ref{eq:phi_component}) that $J=r^2\dot\phi$ is a
constant of the motion.
This might have been predicted from the fact that
$s^{\beta\alpha} f_{[\alpha\mu]} u^\mu$,
the right-hand side of (\ref{eq:modified_geodesic}),
does not have a $\phi$-component,
and as was seen in Section~\ref{sec:conservation_laws}, $J$ is conserved
for geodesic motion.

Contrary to the situation for geodesic and path motion,
$s_{t\beta} u^\beta=\gamma \dot t$
is not conserved.
In fact, we see from (\ref{eq:t_component}) that the actual first integral
of the motion is
\begin{equation}
\label{eq:modified_motion_energy}
E = \gamma \frac{dt}{d\tau}
+ \frac{\lambda\gamma f'}{\sqrt{\alpha\gamma(r^4+f^2)}} .
\end{equation}
That this represents the energy per unit mass can be seen by taking
the large $r$ limit and using (\ref{eq:app_larger_r}).

With these results, we can rewrite (\ref{eq:r_component}) as
\begin{equation}
\label{eq:modified_motion_complete_r}
0 = \frac{d^2r}{d\tau^2}
+ \frac{\alpha'}{2\alpha}\left(\frac{dr}{d\tau}\right)^2
- \frac{J^2}{r^3\alpha}
+ \frac{\lambda}{\alpha\gamma}
\left(E-\frac{\lambda\gamma f'}{\sqrt{\alpha\gamma(r^4+f^2)}}\right)
\frac{d}{dr}\left(\frac{\gamma f'}{\sqrt{\alpha\gamma(r^4+f^2)}}\right)
+ \frac{\gamma'}{2\alpha\gamma^2}
\left(E-\frac{\lambda\gamma f'}{\sqrt{\alpha\gamma(r^4+f^2)}}\right)^2 .
\end{equation}

There are two calculations to perform.
First, we calculate the corrections to the Newtonian
gravitational force
coming from the supplementary forcing term.
Secondly, we will determine the equation for the orbit of a particle
acted upon by this extra force.
In order to extract a useful result from this, we will make certain
simplifying assumptions.
In particular, we will assume that $f(r)/r^2 \ll 1$ and
$\lambda f'(r)/r^2 \ll 1$.
If $s\ll 1$ and $\mu M \ll 1$, this corresponds to regions of
spacetime where $M/r \ll 1$.

Consider first the Newtonian gravitational force.
We rewrite (\ref{eq:modified_motion_complete_r}) in terms of
$r(t)$:
\[
\frac{d^2r}{dt^2} + \frac{\alpha'}{2\alpha}\left(\frac{dr}{dt}\right)^2
- \frac{{J_N}^2}{\alpha r^3} + \frac{\gamma'}{2\alpha}
= - \frac{\lambda\gamma}{\alpha}
\left(E-\frac{\lambda\gamma f'}{\sqrt{\alpha\gamma(r^4+f^2)}}
\right)^{-1}\frac{d}{dr}
\left(\frac{\gamma f'}{\sqrt{\alpha\gamma(r^4+f^2)}}\right) .
\]
Here, $J_N \equiv r^2d\phi/dt$ is the Newtonian value of the
angular momentum per unit mass.
We recognize the left-hand side as the usual GR contributions
to the equation of motion.

To our order of approximation,
\[
\frac{\lambda\gamma}{\alpha}
\left(E-\frac{\lambda\gamma f'}{\sqrt{\alpha\gamma(r^4+f^2)}}
\right)^{-1}\frac{d}{dr}
\left(\frac{\gamma f'}{\sqrt{\alpha\gamma(r^4+f^2)}}\right)
\approx \frac{\lambda}{E}
\frac{d}{dr}\left(\frac{\gamma f'}{\sqrt{\alpha\gamma(r^4+f^2)}}
\right)
\approx \frac{\lambda
s M^2 \mu^2}{3E} \frac{e^{-\mu r}(1+\mu r)}{r^2} .
\]
Therefore, the radial equation of motion may be written
\begin{equation}
\label{eq:look_at_this}
\frac{d^2r}{dt^2} - \frac{{J_N}^2}{r^3} = - \frac{M}{r^2}
- \frac{\lambda
sM^2\mu^2}{3E}\frac{e^{-\mu r}(1+\mu r)}{r^2} ,
\end{equation}
where we have assumed that the particle is moving slowly, so
that $dr/dt \ll 1$.
If $\lambda < 0$, this yields a repulsive Yukawa force
in (\ref{eq:look_at_this}), while $\lambda > 0$ renders this force
attractive.

The Yukawa behaviour generated by this modification of geodesic
motion has interesting phenomenological implications.
Take, for instance, the case of very large $r$.
We find in this case that the right-hand side of (\ref{eq:look_at_this})
reduces to $-M/r^2$; that is, the usual Newtonian law of gravitation.
On the other hand, for $r \ll 1/\mu$ but still with $r \gg M$,
the last term in (\ref{eq:look_at_this}) becomes
\[
- \left(1+\frac{\lambda s M\mu^2}{3E}\right)\frac{M}{r^2} .
\]
Again, this is the usual Newtonian law of gravitation, but with a
renormalized gravitational constant.
Therefore, in the two weak-field regions that we have just considered,
we find that Newtonian gravitation applies.
However, the actual strength of the force is different in each region.
This can be used to fit the galaxy rotation curves without resorting
to postulating dark-matter \cite{bib:Sanders}.

For completeness, we will now derive the equation for the orbit of a
particle in this scheme.
{}From the normalization condition
$s_{\mu\nu}u^\mu u^\nu=1$ and using
$\theta = \pi/2$ and $\dot\theta = 0$, we arrive at
\[
\left(\frac{du}{d\phi}\right)^2
= \frac{1}{\alpha\gamma J^2}
\left(E - \frac{\lambda\gamma f'}{\sqrt{\alpha\gamma(r^4+f^2)}}\right)^2
- \frac{1}{\alpha r^2} - \frac{1}{\alpha J^2} .
\]
Differentiating this with respect to $\phi$ gives the equation
describing the orbit of the particle:
\begin{equation}
\label{eq:orbit_equation}
\frac{d^2u}{d\phi^2}
= - \frac{r^2}{2}\frac{d}{dr}\left[\frac{1}{\alpha\gamma J^2}
\left(E - \frac{\lambda\gamma f'}{\sqrt{\alpha\gamma(r^4+f^2)}}\right)^2
- \frac{1}{\alpha J^2} - \frac{1}{\alpha r^2}\right] .
\end{equation}

Thus far, the treatment has been entirely general.
We can specialize this to the
case where~$f(r)/r^2\ll 1$ and $\lambda f'(r)/r^2 \ll 1$.
Taking the Schwarzschild forms $\gamma = 1/\alpha = 1-2M/r$ and
using (\ref{eq:app_larger_r}), we arrive at the orbit equation:
\begin{equation}
\label{eq:lowest_order_orbit}
\frac{d^2u}{d\phi^2} + u
= \frac{M}{J^2}\left(1
+ \frac{E\lambda sM\mu^2}{12}(1+\mu r)e^{-\mu r}\right) + 3Mu^2 .
\end{equation}
This equation is similar to the orbit equation from
GR (see \cite{bib:Weinberg}, p.~186),
the only difference being
the factor multiplying the angular momentum term on the right-hand side.

A comment should be made about the role played by the quantity
$\lambda$ in these equations.
In a sense, $\lambda$ represents the strength of the coupling between
the test-particle and the NGT field;
it could be thought of as a charge, although unlike the electromagnetic
charge, it does not follow from any sort of flux integral.
The question immediately arises as to whether $\lambda$ is a property
of the test-particle or some universal constant.
If the former, then we must deal with the violations of the
weak equivalence principle that this material-dependence will
cause (see~\cite{bib:Will}, p.~13).
However, since there is no compelling evidence to suggest that the weak
equivalence principle is actually violated in reality \cite{bib:NGT_SEP},
it would seem more sensible to choose the latter route
and take $\lambda$ to be some material-independent constant.

\section{Conclusions}

We have described observational differences between the geodesic
and path equations, in the hope that experiments may differentiate
between the two.
However, despite the differences between these two equations of motion,
we have found that under certain circumstances, for example the case of
radial motion, they predict identical results.
Furthermore, we have shown that in the weak-field limit, both of these
equations lead to virtually identical equations of motion, owing to the
fact that NGT is constructed so as to recover GR in the weak-field limit.

Many of our results are implicitly linked to the case of a static,
spherically symmetric field.
This is a question of practicality: as it stands, this is the only
available solution to the field equations.
However, we would expect that the qualitative effects that we have
described would be generic.
In particular, as pertains to constants of the motion and the like, the
geodesic equation of NGT behaves very much like its counterpart in GR.
This is to be expected: the geodesic equation of NGT is constructed so as
to couple the particle only indirectly to the antisymmetric field.
On the other hand, it was found that the path equation had different
constants of the motion.
In particular, we found that the conventional angular momentum,
$J=\beta\dot\phi\sin^2\theta$, was not a first integral of the motion.
Rather, a different first integral was derived, and was found to be a
sort of ``generalized'' angular momentum.
Although we have no proof to support this, the work we have performed
thus far would lead us to
expect that this generalization of constants of the motion would
be a generic feature, owing to the right-hand side of
(\ref{eq:conserved_momentum}):
whenever we have come across
a case where (\ref{eq:conserved_momentum}) did not yield an
explicit constant of the motion, the resulting equation could
nonetheless be integrated to yield a first integral.

Particularly interesting behaviour was found by studying radial motion
in a static, spherically symmetric field.
Firstly, it was found that an infalling particle need not stop at the
origin $r=0$.
This was seen using the $\nu$-coordinate system, where it was found that
a particle can pass through $\nu_0$ and proceed toward $\nu<\nu_0$.
Furthermore, the particle could reach any radius $\nu_f$ from
any non-zero radius $\nu_i$ in a finite proper time.
An upper bound was placed on the proper time taken to travel this distance.
It is unclear at this time exactly what is implied by this result:
the Wyman solution to the infinite-ranged NGT field equations is a
vacuum solution, devoid of any matter contributions.
As the gravitational collapse problem in NGT remains unsolved,
it is not obvious that the vacuum Wyman solution, as opposed to, say,
the Schwarzschild solution, is in fact the final
state of a stellar collapse.
Regardless, the comparison with the corresponding behaviour in
GR is striking, and should be kept in mind.

The major difference between the geodesic and path equations was found
by studying circular motion.
Here, it was shown that circular orbits for the geodesic equation in NGT
extend to lower radii than in GR, while the last circular orbit for the
path equation lies outside its GR counterpart.
This behaviour was found to be strengthened for larger values of the
parameter $s$ appearing in the static, spherically symmetric field.

After demonstrating that the geodesic and path equations have similar
weak-field limits, we introduced three interactions which could serve
as alternate methods for coupling a test-particle to the antisymmetric
components of the NGT field.
Of these three, only one was found to generate any interaction at all in
a static, spherically symmetric field.
The correction to the Newtonian gravitational force acting on the particle
was worked out in the weak-field limit, and was found to be a Yukawa force.
The correction to the orbit of a particle was also calculated; the
lowest-order NGT correction was shown to be a $1/r^3$ term.

\acknowledgments

This work was supported by the Natural Sciences and Engineering
Research Council of Canada.
J.\ L\'egar\'e would also like to thank the Government of Ontario
for their support during part of this work.
We would like to thank E.\ Demopoulous for his help.
We would also like to thank M.\ Clayton and P.\ Savaria for the countless
hours they spent reading endless copies of this manuscript without losing
patience, and for their many stimulating discussions, many of
which lead to several of the ideas contained in this work.

\appendix

\section{The Static, Spherically Symmetric Solution in NGT}

\label{sec:Wyman_solution}

We list here the salient features and results of the static,
spherically symmetric
solution to the new NGT field equations.
The coordinate system used is $x^0=t$, $x^1=r$, $x^2=\theta$, and
$x^3=\phi$, where $r$, $\theta$, and $\phi$ are the usual spherical
coordinates.

The NGT field equations contain a parameter $1/\mu$ which is identified
with the range of the skew components of the fundamental tensor.
The static, spherically symmetric solution introduces two constants of
integration: $M$ and $s$.
$M$ plays the same role as the Schwarzschild mass in GR, while $s$ is a
dimensionless constant.
Beyond being a measure of the strength of the antisymmetric contributions
to the gravitational field, the physical meaning of $s$ is still unclear.
To be specific, we will assume that $M \ll 1/\mu$.

The fundamental tensor is written
\begin{equation}
\label{eq:app_metric}
g_{\mu\nu} = \left[
\begin{array}{cccc}
\gamma(r) & w(r) & 0 & 0 \\
-w(r) & -\alpha(r) & 0 & 0 \\
0 & 0 & -\beta(r) & f(r)\sin\theta \\
0 & 0 & -f(r)\sin\theta & -\beta\sin^2\theta
\end{array}
\right] .
\end{equation}
In \cite{bib:Clayton} and \cite{bib:Cornish2},
it is shown that the only solution which yields
an asymptotically Minkowskian space has $w(r)=0$.

In the case of infinite-ranged NGT, corresponding to setting $\mu=0$ in
the field equations, an exact solution to the field equations
exists:
\begin{mathletters}
\label{eq:exact_Wyman}
\begin{eqnarray}
\gamma(r) &=& e^\nu \\
\alpha(r) &=& \frac{M^2 e^{-\nu}(1+s^2)}{(\cosh(a\nu)-\cos(b\nu))^2}
\left(\frac{d\nu}{dr}\right)^2 \\
\beta(r) &=& r^2 \\
f(r) &=& \frac{2M^2(\sinh(a\nu)\sin(b\nu)
+ s(1-\cosh(a\nu)\cos(b\nu)))}{e^\nu(\cosh(a\nu)-\cos(b\nu))^2} ,
\end{eqnarray}
\end{mathletters}where
\[
\begin{array}{ccc}
\displaystyle
a = \sqrt{\frac{\sqrt{1+s^2}+1}{2}} &
\hspace{.25in}\mbox{\rm and}\hspace{.25in} &
\displaystyle
b = \sqrt{\frac{\sqrt{1+s^2}-1}{2}} .
\end{array}
\]
The function $\nu(r)$ is determined from the relation
\begin{equation}
\label{eq:really_obvious_relation}
e^{\nu}(\cosh(a\nu)-\cos(b\nu))^2\frac{r^2}{2M^2}
= \cosh(a\nu)\cos(b\nu) - 1 + s\sinh(a\nu)\sin(b\nu) .
\end{equation}
This exact solution is referred to as the Wyman solution
\cite{bib:Wyman,bib:Cornish_Wyman},
and holds for all $s$.

There are two coordinate systems of interest in the Wyman solution.
The first is the conventional set of spherical coordinates used above.
However, (\ref{eq:really_obvious_relation}) can be viewed as defining
a transformation from a different coordinate system, the
$\nu$-coordinates, to conventional spherical coordinates
\cite{bib:nu-coordinates}.
The $\nu$-coordinates are interesting, as (\ref{eq:really_obvious_relation})
need not be inverted; the coordinate system is $x^0=t$, $x^1=\nu$,
$x^2=\theta$ and $x^3=\phi$, where
\[
\alpha(\nu) = \frac{M^2 e^{-\nu}(1+s^2)}{(\cosh(a\nu)-\cos(b\nu))^2} ,
\]
\begin{equation}
\label{eq:beta_in_nu}
\beta(\nu) = \frac{2M^2(\cosh(a\nu)\cos(b\nu)-1+s\sinh(a\nu)\sin(b\nu))}
{e^{\nu}(\cosh(a\nu)-\cos(b\nu))^2} ,
\end{equation}
and with $\gamma(\nu)$ and $f(\nu)$ given as above.
Since (\ref{eq:really_obvious_relation}) gives $r$ as a
many-to-one function of $\nu$, a particular branch of the solution must
be picked.
This selection is done by picking the branch that will yield the positive-mass
Schwarzschild solution as a limit.
This branch begins at $\nu=0$ and extends toward negative $\nu$.
In such a coordinate system, the asymptotic, weak-field region is at
$\nu=0$, while the ``origin'' occurs at $\nu_0$ defined by
$\beta(\nu_0)=0$.
The particular value of $\nu_0$ depends on the value of $s$;
for $s=1$, we find numerically that $\nu_0\approx -5.1667$.

When $f(r)/r^2 \ll 1$, a solution to the linearized field equations is
approximated by
\cite{bib:Cornish2}
\begin{mathletters}
\label{eq:app_larger_r}
\begin{eqnarray}
\gamma(r) &\approx& 1 - \frac{2M}{r} \\
\alpha(r) &\approx& \left(1 - \frac{2M}{r}\right)^{-1} \\
f(r) &\approx& \frac{sM^2}{3}e^{-\mu r}(1+\mu r) ;
\end{eqnarray}
\end{mathletters}the coordinates are the conventional spherical coordinates.
In \cite{bib:Clayton}, a similar result is derived.
The two forms can be seen to be asymptotically equivalent,
notwithstanding a change
in notation, by taking $\mu M \ll 1$ in (2.12) of \cite{bib:Clayton} and
taking $\mu r \gg 1$ in the above result.

For both conventional spherical coordinates and the $\nu$-coordinates,
the Christoffel symbols are given by:
\begin{mathletters}
\label{eq:Christoffels}
\begin{eqnarray}
\Christoffel{t}{tx}
&=& \frac{\gamma'}{2\gamma} \\
\Christoffel{x}{tt}
&=& \frac{\gamma'}{2\alpha} \\
\Christoffel{x}{xx}
&=& \frac{\alpha'}{2\alpha} \\
\Christoffel{x}{\phi\phi}
&=& \sin^2\theta\Christoffel{x}{\theta\theta}
= -\frac{\beta'\sin^2\theta}{2\alpha} \\
\Christoffel{\theta}{x \theta}
&=& \Christoffel{\phi}{x \phi}
= \frac{\beta'}{2\beta} \\
\Christoffel{\theta}{\phi\phi}
&=& -\sin\theta\cos\theta \\
\Christoffel{\phi}{\theta\phi}
&=& \frac{\cos\theta}{\sin\theta} , \\
\end{eqnarray}
\end{mathletters}where $x$ represents either $r$ or $\nu$.
A prime denotes differentiation with respect to $x$.

The $\Gamma$-connection coefficients,
$\Gamma^\beta_{\mu\nu}$, are determined
by solving the NGT compatibility condition:
\begin{equation}
\label{eq:app_compatibility}
\partial_\eta g_{\lambda\xi} - g_{\rho\xi}\Gamma^\rho_{\lambda\eta}
- g_{\lambda\rho}\Gamma^\rho_{\eta\xi}
= \frac{1}{6}g^{(\mu\rho)}(g_{\rho\xi}g_{\lambda\eta}
- g_{\eta\xi}g_{\lambda\rho} - g_{\lambda\xi}g_{[\rho\eta]})W_\mu ,
\end{equation}
where $W_\mu$ is determined from
\[
\partial_\rho{\bf g}^{[\sigma\rho]}
= -\frac{1}{2}{\bf g}^{(\rho\sigma)} W_\rho .
\]
When $w(r)=0$, it can be shown that $W_\mu=0$.
The right-hand side of (\ref{eq:app_compatibility}) therefore
vanishes.
For both conventional spherical coordinates and the $\nu$-coordinates,
the nonvanishing components of the $\Gamma$-connection are:
\begin{mathletters}
\label{eq:app_gammas}
\begin{eqnarray}
\Gamma^t_{(tx)} &=& \frac{\gamma'}{2\gamma} \\
\Gamma^x_{(tt)} &=& \frac{\gamma'}{2\alpha} \\
\Gamma^x_{(xx)} &=& \frac{\alpha'}{2\alpha} \\
\Gamma^x_{(\phi\phi)} &=& \sin^2\theta\Gamma^x_{(\theta\theta)}
= -
\frac{(\beta'(\beta^2-f^2)+2ff'\beta)\sin^2\theta}{2\alpha(\beta^2+f^2)} \\
\Gamma^\theta_{(x\theta)} &=& \Gamma^\phi_{(x\phi)}
= \frac{\beta\beta'+ff'}{2(\beta^2+f^2)} \\
\Gamma^\theta_{(\phi\phi)} &=& - \sin\theta\cos\theta \\
\Gamma^\phi_{(\theta\phi)} &=& \frac{\sin\theta}{\cos\theta} \\
\Gamma^x_{[\theta\phi]} &=&
\frac{(f'(\beta^2-f^2)-2\beta\beta'f)\sin\theta}{2\alpha(\beta^2+f^2)} \\
\Gamma^\theta_{[x\phi]} &=& - \sin^2\theta\Gamma^\phi_{[x\theta]}
= \frac{(\beta f'-\beta'f)\sin\theta}{2(\beta^2+f^2)} .
\end{eqnarray}
\end{mathletters}Again, $x$ denotes either $r$ or $\nu$, and
a prime denotes differentiation with respect to $x$.


\begin{references}
\bibitem{bib:nu-coordinates} Note that, historically, it is the
$\nu$-coordinates which were used to obtain a solution to the
field equations of old NGT or Einstein-Straus Unified Field Theory.
We have found it more convenient here to approach things from the
opposite direction and to view (\ref{eq:really_obvious_relation}) as
defining the $\nu$-coordinate system in terms of conventional spherical
coordinates.
\end{references}
\end{document}